\documentstyle[masaps,graphics]{revtex}
\def\p{^\prime}
\def\h#1{\hat{#1}}

\def\gam{\gamma}
\def\tr{\hbox{\large Tr\,}}
\def\k{{\bf k}}
\def\kp{{\bf k^\prime}}
\def\x{{{\bf x}_1}}
\def\xx{{{\bf x}_2}}
\def\xp{{{\bf x}^\prime_1}}
\def\xxp{{{\bf x}^\prime_2}}
\def\r{{{\bf r}_1}}
\def\rr{{{\bf r}_2}}

\def\<{\langle}
\def\>{\rangle}
\def\tpc{(2\pi)^3}
\def\tpone{\hbox{\vbox{%
\hbox to 10pt{\hfil$\otimes$\hfil}%
\hrule height 1pt width 0pt%
\hbox to 10pt{\hfil\tiny I\hfil}%
\hrule height -5pt depth 5pt}}}
\def\tptwo{\hbox{\vbox{%
\hbox to 10pt{\hfil$\otimes$\hfil}%
\hrule height 1pt width 0pt%
\hbox to 10pt{\hfil\tiny II\hfil}%
\hrule height -5pt depth 5pt}}}
\def\tpthree{\hbox{\vbox{%
\hbox to 10pt{\hfil$\otimes$\hfil}%
\hrule height 1pt width 0pt%
\hbox to 10pt{\hfil\tiny III\hfil}%
\hrule height -5pt depth 5pt}}}

\title{Tensor product expansions for correlation in quantum many--body systems}
\author{G\'abor Cs\'anyi$^\dag$ and T.A. Arias$^\ddag$\\
$\dag$ Department of Physics, Massachusetts Institute of Technology, Cambridge, MA 02139\\
$\ddag$ Laboratory of Atomic and Solid State Physics, Cornell
University, Ithaca, NY 14853\\
}

\twocolumn

\abstract{
We explore a new class of computationally feasible
approximations of the two--body density matrix as a finite sum of
tensor products of single--particle operators.  Physical symmetries
then uniquely determine the two--body matrix in terms of the one--body
matrix.  Representing dynamical correlation alone as a single tensor
product results in a theory which predicts near zero dynamical
correlation in the homogeneous electron gas at moderate to high
densities.  But, representing both dynamical and statistical
correlation effects together as a tensor product leads to the recently
proposed ``natural orbital functional.''  We find that this latter
theory has some asymptotic properties consistent with established
many--body theory but is no more accurate than Hartee--Fock in
describing the homogeneous electron gas for the range of densities
typically found in the valence regions of solids.
\break
\vrule height 15pt width 0pt
PACS 71.10.-w 71.15.Mb, Accepted for publication in Physical Review B}

\begin{document}
\maketitle

\bigskip

The fundamental difficulties associated with {\em ab initio} solutions
of the quantum mechanical many--body problem stem directly from the
large dimensionality of the wave function.  Rather than dealing
directly with the many--body wave function, large scale electronic
structure calculations determine the ground--state energy by minimizing
an energy functional over the much more manageable space of
single--particle orbitals.  Hohenberg--Kohn--Sham theory
\cite{HohenbergKohn,KohnSham} places this approach on a firm
theoretical footing by establishing that exact electronic ground-state
energies can be determined in this manner by minimizing a universal
functional of such orbitals.

Although the requisite universal energy functional is not known
exactly, the local density approximation\cite{KohnSham} and various
improvements thereon
\cite{Goedecker,PerdewYue,JuanKaxiras,Ortiz,Becke} are sufficient to
resolve bond energies to a little better than one--tenth of an
electron Volt.  With this accuracy, these functionals make possible
highly predictive first principles studies in diverse areas such as
the study of surfaces, point defects, plastic deformation and chemical
reactions.  (For a review see\cite{dftreview}.)  Although sufficient
for many studies, this error is too large to allow accurate prediction
of the rates of microscopic processes at room temperature, thus
limiting the ultimate predictive power of the {\em ab initio}
density--functional approach.  Whereas further improvement of energy
functionals for single--particle theories is an active area of
research, much less work has been done to construct energy functionals
of the one--body density matrix\cite{Goedecker,Davidson}.  Such
functionals are promising because the density matrix contains more
explicit information than do the Kohn--Sham orbitals, and an accurate
energy functional thus is likely to be simpler in form.  The exact
kinetic energy functional, for instance, is known for the density
matrix, but not the Kohn--Sham orbitals.

To date, the only extensively used density--matrix energy functional is
the Hartree--Fock approximation.  One may view Hartree--Fock theory as
{\em approximating} the two--body matrix as a sum of Hartree and
exchange terms, each of which is a tensor product of the one--body
density matrix.  Hartree--Fock then uses the known, {\em exact} energy
functional of the two--body density matrix to produce an energy
functional of the one--body density matrix.  In this work, we build
upon this perspective and consider tensor product approximations to the
two--body density matrix which not only serve as generators of energy
functionals of the one--body density matrix but also provide estimates
of the two--body matrix in terms of the one--body matrix and thereby
shed light on the nature of correlation in quantum many--body systems.

{\em Notation ---} For a system of $N$ electrons, the exact energy
functional of the two--body density matrix $\gam(\x\xx,\xp\xxp) \equiv
\langle \Psi | \h\psi^\dagger(\xp) \h\psi^\dagger(\xxp) \h\psi(\xx)
\h\psi(\x) | \Psi \rangle/2$, 
is
\begin{eqnarray}
E & = & \int d\x \left. \left(\left[-\frac12\nabla^2_\r 
+U(\r)\right]
n(\x, \xp)\right)\right|_{\xp=\x} \label{totE}\\
&&+ \int d\x d\xx\, \frac{\gam(\x\xx,\x\xx)}{|\r-\rr|}. \nonumber 
\end{eqnarray}
Here and throughout, we work in atomic units, and the ${\bf x}_i$ are
compound coordinates representing both position ${\bf r}_i$ and spin
$s_i$ so that integration over ${\bf x}_i$ represents integration over
space and summation over spin channels.  The external potential and
one--body density matrix are $U({\bf r})$ and $n(\x, \xp) \equiv
\langle \Psi | \h\psi^\dagger(\xp) \h\psi(\x) | \Psi \rangle$,
respectively.  Finally, the one--body density matrix comes directly
from $\gam$ through the sum--rule
\begin{equation}
\left(\frac{N-1}2\right) n(\x,\xp) = \int\!d\xx\,\gam(\x\xx,\xp\xx).
\label{sumrule}
\end{equation}

{\em Computational considerations ---} Although (\ref{totE}) is exact,
the two--body density matrix $\gam$ is a function of four variables, and
direct computation with such functions is infeasible.  Viable
techniques, however, exist for dealing directly with two--variable
functions such as the one--body density
matrix\cite{MGC,LNV,Daw,ODMG,HG,K}.  Accordingly, we expand $\gam$ in
terms of two--variable functions,
\begin{equation}
\h\gam = \sum_i \,\, \h g_i\otimes\h h_i,
\label{gamexpansion}
\end{equation}
which, by separation of variables, is always possible.  Here, $\h g_i$
and $\h h_i$ are one--body operators (functions of two variables), we
view the two--body density matrix as a two--body operator (a function of
four variables), and the tensor product denotes one of the three
possible choices for separating four variables,
\begin{eqnarray*}
\h g\tpone \h h &=& g(\x,\xp)h(\xx,\xxp)\\
\h g\tptwo \h h &=& g(\x,\xx)h(\xxp,\xp)\\
\h g\tpthree \h h &=& g(\x,\xxp)h(\xx,\xp).
\end{eqnarray*}

{\em Quantum statistical considerations ---} We wish to maximize the
physical content of a truncation of the expansion (\ref{gamexpansion})
to a limited number of terms.  Physically, $n(\x,\xp)$ is the quantum
amplitude for the insertion of a hole at position $\x$ and its {\em
instantaneous} removal from $\xp$, whereas $\gam(\x\xx,\xp\xxp)$ is
the quantum amplitude for the insertion of a pair of holes at $\x,
\xx$ and their removal from $\xp, \xxp$.  Approximating the latter
events as independent gives $\h\gam \approx \h\gam_H \equiv (\h n
\tpone \h n)/2$, the familiar Hartree approximation, where the factor
of two maintains the normalization implicit in (\ref{sumrule}).  We
may then refine this mean--field behavior and define $\h\gam \equiv
\h\gam_H+\h\gam_{xc}$, where $\h\gamma_{xc}$ represents exchange and
correlation effects which we then expand, without loss of generality,
according to (\ref{gamexpansion}).

The most significant drawback of the Hartree approximation is that
$\gam_H$ is not properly antisymmetric with respect to particle
exchange ($\xp \leftrightarrow \xxp$), so that, potentially, many
terms will be required in $\h\gamma_{xc}$ to restore this symmetry.
Alternately, we may explicitly ensure antisymmetry and take $\h\gam =
\h\gam_{HF} + \h\gam_c \equiv (\h n \tpone \h n - \h n \tpthree \h
n)/2 + \h\gam_c$.  In this form, $\h\gam_{HF}$ is precisely the
familiar Hartree--Fock approximation.  Again, we may expand the
unknown $\h\gam_c$ as in (\ref{gamexpansion}).

Finally, we note that a fundamental difficulty exists when using a
finite number of terms of Type II.  Because of the instantaneous
nature of the quantum event which $\gam$ represents, we expect for
normal systems that $\gam \rightarrow 0$ as the insertion $\xp\xxp$
and removal $\x\xx$ locations are placed at ever further distances
from one another.  In an extended system with the expansion for $\gam$
containing a finite sum of terms of Type II, taking this limit while
keeping the insertion points near one another and the removal points
near one another will violate this asymptotic condition unless each
Type II term is identically zero.  Accordingly, we focus below on
approximating $\h\gamma_{xc}$ and $\h\gamma_{c}$ with terms of Type I
and III.

{\em Symmetry --- } Physical symmetries of $\h\gam$ significantly
restrict the allowable forms for the tensor products appearing in
expansions of the form (\ref{gamexpansion}) for $\h\gamma_{xc}$ and
$\h\gamma_{c}$.  We first consider Hermiticity $\gam(\x\xx,\xp\xxp) =
\gam^*(\xp\xxp,\x\xx)$ and particle permutation symmetry
$\gam(\x\xx,\xp\xxp) = \gam(\xx\x,\xxp\xp)$ for each of the three
types of tensor product and defer discussion of Fermionic antisymmetry
until after the sum--rule immediately below.  Because both $\gam_H$
and $\gam_{HF}$ respect these symmetries, so must tensor products
representing $\h\gamma_{xc}$ and $\h\gamma_{c}$.  These symmetries
imply relations with arbitrary constants of proportionality between
the one--body operators of such products which may be absorbed into
the one--body operators. Table~\ref{symmtable} {\em summarizes} the
final result of such considerations for all three tensor products.  We
note that in all cases symmetry restricts the tensor product to
involve only a {\em single} one--body operator which remains to be
determined.

{\em Sum--rule --- } Remarkably, when a single term satisfying
Hermiticity and particle permutation is used to represent
$\h\gamma_{xc}$ or $\h\gamma_{c}$, we can ``invert'' the sum--rule
(\ref{sumrule}) and determine the two--body density matrix $\gam$ in
terms of the one--body matrix $n$.  Inserting the resulting form for
$\gam$ into (\ref{totE}) then generates an energy functional of the
one--body matrix which implicitly satisfies the sum--rule, an
important property for obtaining good ground state energies.

When terms of Type I represent $\h\gam_{xc}$ or $\h\gam_{c}$, the
sum rule becomes $\h g (\tr \h g) =\h o$, where $\h o =$ $\h n$ or $\h
n (1-\h n)$, respectively.  The solution of this equation is $\h g =
\pm \h o/\sqrt{\tr \h o}$.  In either case, the extensivity of $\tr \h
o$ makes the resulting solution irrelevant in the thermodynamic limit
$N\rightarrow \infty$.  For products of Type III, the sum--rule
combined with appropriate symmetries gives $\h g^2 =\h o$, so that $\h
g = \sqrt{\h o}$ with $\h o$ defined as above.

{\em Fermionic antisymmetry ---} Fermionic antisymmetry,
$\gam(\x\xx,\xp\xxp) = - \gam(\x\xx,\xxp\xp)$, is a stronger
condition than the particle permutation symmetry condition considered
above and imposes more complex constraints.  Two--body density
matrices $\gam$ which satisfy this condition may always be written
with Type I and III products appearing in corresponding pairs of the
form $\h g \tpone \h g - \h g \tpthree \h g$ and with Type II products
(were we to consider such)
whose one--body terms are separately antisymmetric.

The two viable functionals remaining after the above considerations
involve a Type III representation for either $\h\gam_{xc}$ or
$\h\gam_{c}$.  Unfortunately these forms do not consist of symmetric
pairs of Type I and Type III products.  To satisfy antisymmetry, one
could take $\h\gam_{xc} = - (\h n \tpthree \h n)/2$, so that $\h\gam=
(\h n \tpone \h n - \h n \tpthree \h n)/2$, which is simply the
Hartree--Fock approximation.  To go beyond this, antisymmetry requires
representing $\h\gam_c$ as a pair of terms so that $\h\gam= (\h n
\tpone \h n - \h n \tpthree \h n + \h g \tpone \h g - \h g \tpthree \h
g)/2$.

The sum--rule for this latter extension on Hartree--Fock is $\h g^2 -\h
g (\tr \h g) = \h n (1-\h n)$, which has no solution for $\h g$ in
extended systems unless $\tr \h g$ vanishes in the thermodynamic
limit.  Under this condition, we have the solution $\h g = \pm \sqrt{
\h n(1-\h n) }$, where the signs of the eigenvalues in the square root
must be chosen to ensure $\tr \h g=0$.  For the paramagnetic phase we
consider below, the natural choice is to take the upper sign for the
spin--up block of the density matrix and the lower sign for the
spin--down block.  The structure of the energy functional
(\ref{totE}), however, is such that for this choice the resulting
energy functional of $\h n$ is equal to that generated by representing
$\h\gam_{c}$ as a single Type III product.

{\em Representation of the functionals ---} The above discussion
leaves the energy of two representations of the two--body matrix to
explore: corrected Hartree theory, $\h\gam = \h\gam_{H} - (\sqrt{\h n}
\tpthree \sqrt{\h n})/2$, and corrected Hartree--Fock theory, $\h\gam
= \h\gam_{HF} - (\sqrt{\h n(1-\h n)} \tpthree \sqrt{\h n(1-\h n)})/2$.
The corresponding energy functionals may be represented either
directly in terms of the one--body density matrix for use with direct
density-matrix methods\cite{MGC,LNV,Daw,ODMG,HG,K}, or in terms of the
spectral (``natural orbital'') representation of the density matrix.

The energy functional (\ref{totE}) contains one--body terms which may
be evaluated directly in terms of the one--body density matrix.  The
remaining two--body term may be computed in terms of the two--point
density $c(\r,\rr) \equiv \sum_{s_1 s_2} \gam(\x\xx,\x\xx)$.  When the
above functionals represent paramagnetic states, this density has the
following form in terms of the eigenvectors (natural orbitals)
$\phi_i$ and corresponding eigenvalues (occupancies) $f_i$ of the
one--body density matrix $\hat n$,
\begin{eqnarray}
c(\r,\rr) & = & {1\over2}\left[
4 \sum_{ij} f_i f_j\phi_i(\r)\phi_i^*(\r)\phi_j(\rr)\phi_j^*(\rr)
\label{goedecker} \right. \\
&& \left. - 2 \sum_{ij} g_{ij}\phi_i(\r)\phi_i^*(\rr)\phi^*_j(\r)\phi_j(\rr)
\right], \nonumber
\end{eqnarray}
where for corrected Hartree and Hartree--Fock theories, we take
$g_{ij}=\sqrt{f_i f_j}$ and $g_{ij}=f_i f_j + \sqrt{f_i(1-f_i)
f_j(1-f_j)}$, respectively.

The first of the above forms corresponds to the natural orbital
functional proposed by Goedecker and Umrigar \cite{Goedecker} as an
{\em ansatz} among the many possible choices for the $g_{ij}$ which
satisfy the sum--rule.  Here, we see that this form is the unique Type
III correction to Hartree theory which satisfies basic symmetries and
the sum--rule.  This theory has been studied analytically in the case
of the homogeneous electron gas in the low density regime ($r_s >
5.77$)\cite{cio}.  Here, we present results for both this theory and
our new corrected Hartree-Fock theory over a range of densities which
includes both this low density regime and the regime more physically
relevant in the valence region of solids, $r_s < 5.77$.

{\em Homogeneous electron gas ---} Translational invariance in the
homogeneous gas ensures that the natural orbitals are plane waves.
The remaining freedom in $\h n$ lies in the occupancy eigenvalues
$f_\k$.  For the paramagnetic phase, the energy (\ref{totE}) then
takes the following form,
\begin{eqnarray}
E & = & 2\int\!{Vd\k\over\tpc}{{\k^2\over2}f_\k} - {1\over V}\int\!{V^2 d\k d\kp \over (2\pi)^6} g_{\k \kp}
{4\pi\over|\k\p-\k|^2}, \label{efunctional} 
\end{eqnarray}
where $V$ is the volume of the system and the factor of two in the
kinetic energy arises from the sum over spin.

The minimum of this energy functional occurs when ${\delta (E-\mu
N)/\delta f_\k} = 0$.  For both corrected Hartree and corrected
Hartree--Fock, this gives to leading order as $|k|\rightarrow\infty$,
$$
{\delta (E-\mu N)/\delta f_\k} \sim k^4 - C/\sqrt{f_\k},
$$
where $C$ is a constant.  Hence, for large $k$, the momentum density
scales as $f_\k \sim |\k|^{-8}$, in agreement with the random phase
approximation and other many--body calculations\cite{Farid}.  This
analysis holds for the present two functionals at all particle number
densities and, in fact, for all theories of the form of
(\ref{efunctional}) in which $g_{\k\kp}\rightarrow\sqrt{f_\k f_\kp}$
as $k\rightarrow\infty$ and $f_\k\rightarrow 0$.

{\em Numerical Results ---} To determine the total energy of the
homogeneous electron gas within the corrected Hartree and
Hartree--Fock approximations, we exploit spherical symmetry and reduce
$f_\k$ to a single variable function $f(k)$, which we represent on a
radial mesh of variable spacing.  We then minimize (\ref{efunctional})
with conjugate--gradients techniques subject to both the constraint that
$f(k)$ corresponds to a total of $N$ electrons and the Fermi constraint that
$0\leq f(k)\leq1$.  The latter constraint we implement by defining
$f(k) \equiv (e^{-x}+1)^{-2}$, where $x$ is then free to range over the real line.

To evaluate the energy functional in terms of the fillings, we
integrate numerically the kinetic and exchange--correlation terms in
(\ref{efunctional}).  A key step in producing accurate results is to
avoid numerical integration across the familiar logarithmic
singularity at $k=k\p$.  For corrected Hartree theory, we avoided this
by integrating the exchange--correlation terms once by parts
analytically and then evaluating the result numerically.  For the
corrected Hartree--Fock case, we found it more efficient to
approximate the Coulomb potential by a Yukawa potential, $e^{-\kappa
r}/r$, and numerically extrapolate the results to $\kappa=0$.  For
these latter calculations, we found it more efficacious to work
in the grand canonical ensemble (holding the chemical potential fixed
while minimizing $E-\mu N$) than to impose the particle--number
constraint explicitly.

The above procedures introduce four sources of numerical error: the
finite spacing of the radial mesh, the finite extent of the mesh in
$k$--space, the error in the extrapolation in the case of using the
Yukawa potential, and the finite number of iterations involved in our
searches for the minima.  Through analysis and numerical experiments,
we determine these errors to be less than $10^{-3}$, $10^{-4}$,
$10^{-3}$ and $10^{-6}$ Hartree per particle, respectively, over the
range of densities which we explored ($0.01 < r_s < 30$).

We found the minimization of the above functionals problematic for two
reasons.  First and foremost, the discretized representation of the
one--dimensional function $f(k)$ on the radial mesh artificially
stabilizes any discontinuities in $f(k)$ which may arise during the
minimization process.  This may be managed by taking care to ensure
that the radial mesh provides sufficient resolution to describe any
large gradients which arise in $f(k)$.  An additional factor making
the calculations difficult is that both the corrected Hartree and
corrected Hartree--Fock functionals suffer from a very large condition
number (in excess of $10^7$).

\begin{figure}
\scalebox{0.4}{\includegraphics{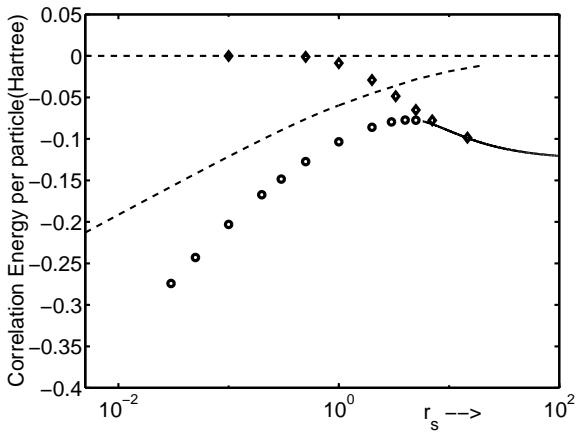}}
\medskip
\caption{Correlation energy per particle of the uniform electron gas.
Hartree--Fock (horizontal dashed line); exact many-body
results\protect\cite{Mahan,CPA} (dashed curve), corrected Hartree
theory (open circles), corrected Hartree--Fock theory (diamonds),
analytic results\protect\cite{cio} for corrected Hartree theory in the low
density limit (solid curve).}
\label{fig:correner}
\end{figure}

Figure~\ref{fig:correner} compares our results for the correlation
energy per particle of the electron gas with established many--body
predictions.  We find that at very low densities, both new theories
begin to approach one another as $r_s \rightarrow \infty$.  For even
moderately low densities ($r_s > 6$), both theories perform
significantly worse than Hartree--Fock, overestimating the correlation
energy by factors from two to four and higher.  For the densities
experienced in the valence regions of solids ($6 > r_s > 1$),
corrected Hartree--Fock theory (diamonds) crosses over from
over--correlation to under--correlation, whereas corrected Hartree
theory (circles) over--correlates by roughly the same amount by which
traditional Hartree--Fock theory under--correlates.  For higher
densities more representative of atomic cores ($1 > r_s$), the
corrected Hartree--Fock theory predicts near zero correlation.  At
these same densities, corrected Hartree theory remains
over--correlated and gives a somewhat better estimate of the
correlation energy than does traditional Hartree--Fock.  The
asymptotic form of the corrected Hartree results in the high density
limit appears to be a straight line in our plot, which corresponds to
the form of the leading order terms of the Gell--Mann--Brueckner
expansion\cite{GMB}.  However, the prefactors which we find by a least
squares fit, $E \approx 0.0570 \log(r_s)-0.0714$, are only correct to
within a little better than a factor of two: the well--known analytic
values for these coefficients are $0.0311$ and $-0.0480$,
respectively.  This suggests that the favorable reports for atoms for
the natural orbital functional\cite{Goedecker} may be in part a
consequence of the high densities experienced in atomic cores, the
limited particle number in the low density regions, and the
self--interaction corrected nature of those calculations.

\begin{figure}
\centerline{\scalebox{0.4}{\includegraphics{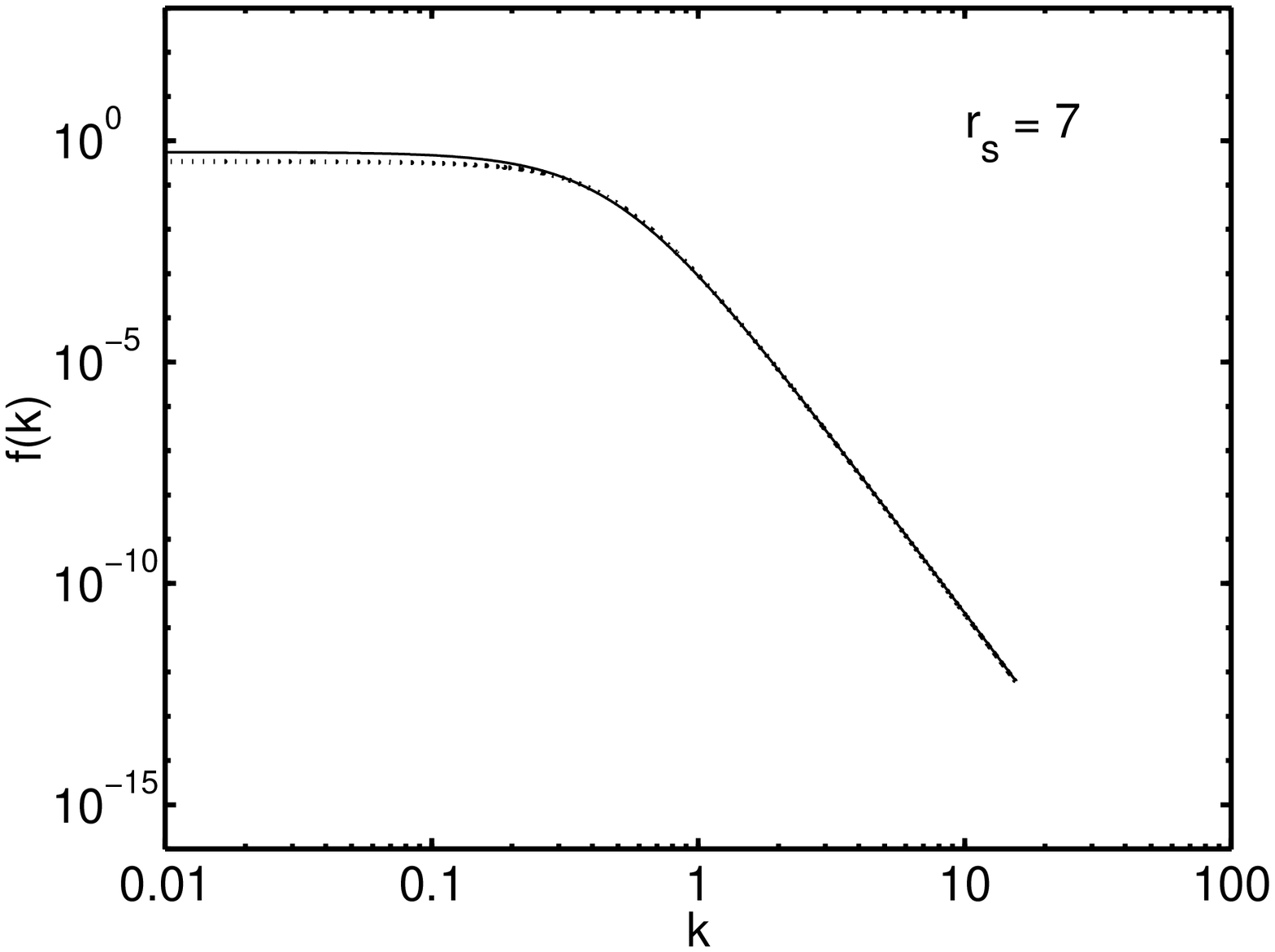}}}
\medskip
\centerline{\scalebox{0.45}{\includegraphics{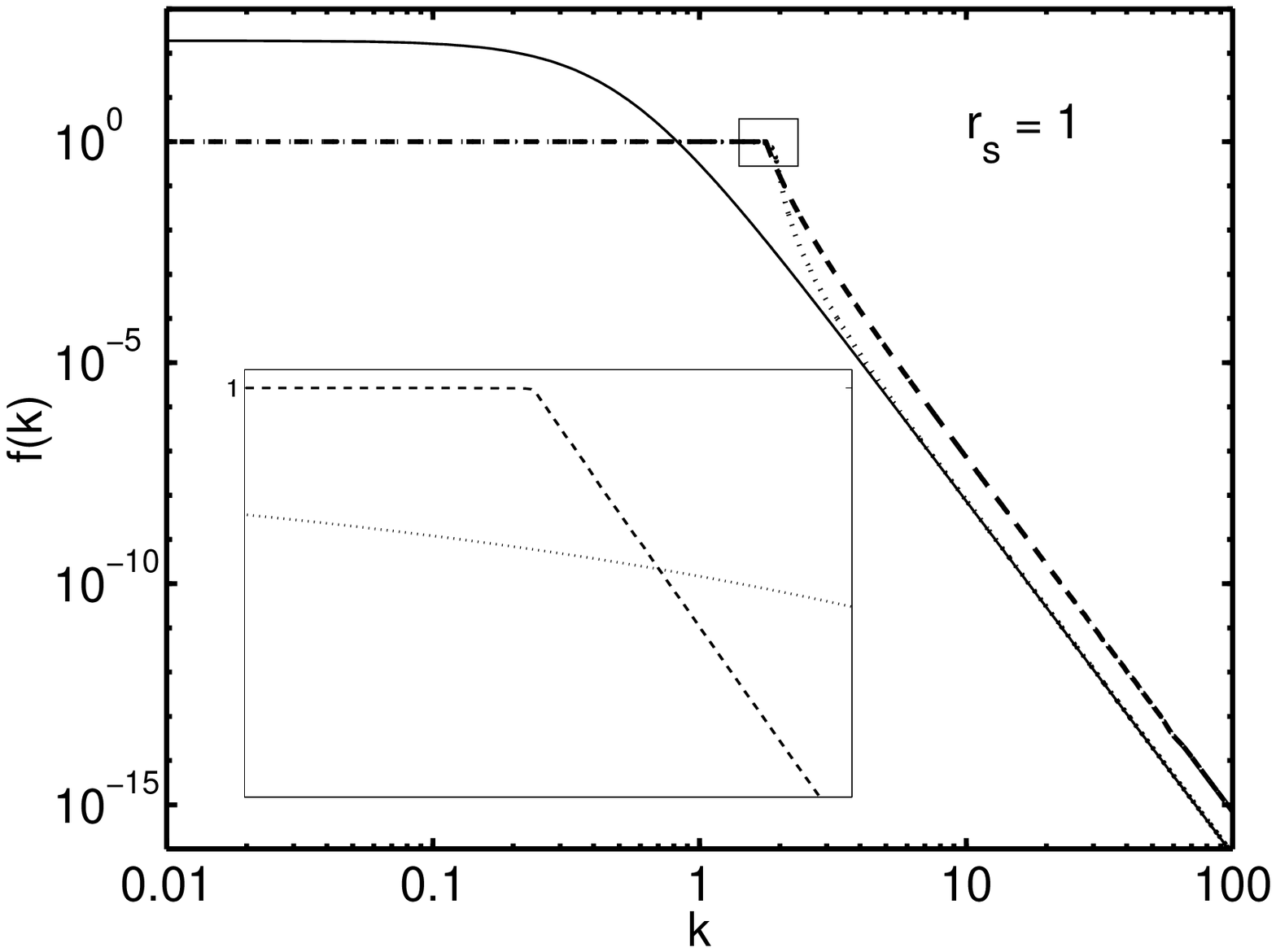}}}
\medskip
\caption{Predictions of momentum distribution of the uniform electron
gas for $r_s = 7$ and $r_s=1$: corrected Hartree theory (dashed
curves), corrected Hartree--Fock theory (dotted curves), analytic
results\protect\cite{cio} for corrected Hartree theory in the low
density limit (solid curve).  Inset shows magnified view.}
\label{fig:momdistr}
\end{figure}

Figure~\ref{fig:momdistr} shows momentum distributions for $r_s = 7$
and $r_s = 1$, which are representative of what we find at low and
high densities, respectively.  We note that the numerical results of
both theories exhibit the correct $k^{-8}$ scaling at large momenta
and that we have very good agreement with the analytic results for the
natural orbital functional\cite{cio} (applicable for $r_s > 5.77$).

We observe that at low densities the momentum distributions of both
corrected Hartree theory and corrected Hartree--Fock theory resemble
one another quite closely and, in fact, approach one another as $r_s$
approaches infinity.  This occurs because in this limit both theories
predict very low occupations $f_i << 1$ so that the central quantities
in the two theories approach one another, $f_i f_j+\sqrt{f_i(1-f_i)
f_j (1-f_j)} \rightarrow \sqrt{f_i f_j}$.

The next panel of the figure illustrates that for densities outside
the range of the applicability of the analytic solution ($r_s <
5.77$), this solution predicts $f>1$ for sufficiently low momenta.  We
find that for such densities, the momentum distributions of corrected
Hartree and corrected Hartree-Fock theory exhibit different behaviors
at low momenta.  Our numerical results indicate (to better than $1$
part in $10^5$) that the momentum distribution of corrected Hartree
theory saturates to $f=1$ near the Fermi level.  (See inset.)  Such
behavior is in direct contradiction of analytic and numerical
many--body results for the electron gas\cite{Farid,CPA} and is
contrary to the previously reported experience with atoms for this
functional\cite{Goedecker}.  In contrast, the momentum distribution of
corrected Hartree--Fock theory never saturates the Fermi occupation
constraint, but rather approaches a value at $k=0$ which remains less
than unity.  We find that in the high density limit the corrected
Hartree--Fock momentum distribution in fact approaches the standard
Hartree--Fock result, ultimately leading to zero correlation energy
for this theory in the high density limit.

In conclusion, we find that when the two--body density matrix is
expanded beyond traditional Hartree and Hartree--Fock theories in
terms of tensor products of one--body operators, fundamental
symmetries allow the exchange--correlation hole sum--rule to be
inverted to give estimates for the two--body matrix, and thus
correlations, in terms of the one--body matrix.  By approximating the
two--body density matrix in terms of the one--body density matrix,
this approach also leads to new energy functionals of the one--body
density matrix.

The addition of the first separable term beyond Hartree theory shows
some promising features when applied to the homogeneous electron gas,
such as giving the correct scaling behavior of the momentum
distribution in the high momentum limit and the correct scaling
behavior of the correlation energy in the high density limit.
However, the electronic states predicted by this functional are
significantly {\em over}--correlated, and attempts to improve upon it
by the addition of further terms without additional constraints would
only add more freedom in the minimization and lead to further
over--correlation.  Hence, before adding a second term, it is
important to use antisymmetry to constrain the first such term to be
the traditional Fock term.  This gives the second functional which we
considered, corrected Hartree--Fock theory.  This second theory
under--correlates the electron gas in the high density limit but still
over--correlates at lower densities.  This indicates that the addition
of further terms will only hold the promise of producing a more
accurate functional if additional, appropriate constraints are imposed.

\begin{acknowledgements}
CSG would like to thank Sohrab Ismail--Beigi for very useful
discussions.  This work was supported primarily by the MRSEC Program
of the National Science Foundation under award number DMR 94--00334.

\end{acknowledgements}

\def\tablestrut{\vrule height 10pt depth 7pt width 0pt}
\def\vertline{\vrule width 0.2pt}
\begin{table}
\begin{tabular}{lccc}
\tablestrut&&Hermiticity&Particle Permutation\\ \hline
\tablestrut$\h g\tpone \h h$&\vertline&$\h g=\h g^\dagger$, $\h h= \h h^\dagger$&$\h g=\h h$\\
\tablestrut$\h g\tptwo\h h$&\vertline&$\h g=\h h^\dagger$&$\h g=\pm \h g^{\rm T}$, $\h h=\pm \h h^{\rm T}$\\
\tablestrut$\h g\tpthree \h h$&\vertline&$\h g=\pm\h h^\dagger$&$\h g = \h h$\\
\end{tabular}
\smallskip
\caption{Implications of Hermiticity and particle permutation
symmetries for tensor product
expansions of $\gam$.}
\label{symmtable}
\end{table}

\end{document}